\documentclass[a4paper,fleqn,usenatbib,useAMS]{mnras}

\usepackage{tikz,xcolor,hyperref}

\definecolor{lime}{HTML}{A6CE39}
\DeclareRobustCommand{\orcidicon}{
	\begin{tikzpicture}
	\draw[lime, fill=lime] (0,0) 
	circle [radius=0.13] 
	node[white] {{\fontfamily{qag}\selectfont \tiny ID}};
	\draw[white, fill=white] (-0.0625,0.095) 
	circle [radius=0.007];
	\end{tikzpicture}
	\hspace{-2mm}
}

\foreach \x in {A, ..., Z}{\expandafter\xdef\csname orcid\x\endcsname{\noexpand\href{https://orcid.org/\csname orcidauthor\x\endcsname}
			{\noexpand\orcidicon}}
}

\usepackage{graphicx}	
\usepackage{amsmath}	
\usepackage{amssymb}	
\usepackage{multicol}        
\usepackage{bm}		
\usepackage{pdflscape}	
\usepackage[encapsulated]{CJK}
\usepackage{ucs}
\usepackage[utf8x]{inputenc}
\usepackage{multirow}
\usepackage{booktabs,caption}
\usepackage[flushleft]{threeparttable}

\usepackage[T1]{fontenc}
\usepackage{ae,aecompl}

\usepackage{newtxtext,newtxmath}

\title[IFS Study of QU\,Vul]{QU\,Vul: An Integral Field Spectroscopy Case Study of a Nova Shell}

\author[Santamar\'\i a et al.]
{E.\ Santamar\'{i}a$^{1,2\thanks{E-mail: {\bf edgar.santamaria8808@alumnos.udg.mx}}\orcidA{}}$,
M.A.\,Guerrero$^{3\orcidB{}}$,
J.A.\,Toal\'{a}$^{4\orcidC{}}$,
G.\,Ramos-Larios$^{1,2\orcidD{}}$
and L.\,Sabin$^{5\orcidE{}}$
\\
$^1$Universidad de Guadalajara, CUCEI, Blvd. Marcelino Garc\'\i a Barrag\'an 1421, 
44430, Guadalajara, Jalisco, Mexico \\
$^2$Instituto de Astronom\'\i a y Meteorolog\'\i a, Dpto.\ de F\'\i sica,
CUCEI, Av.\ Vallarta 2602, 44130, Guadalajara, Jalisco, Mexico\\
$^3$Instituto de Astrof\'\i sica de Andaluc\'\i a, IAA-CSIC, Glorieta de la
Astronom\'\i a s/n, 18008, Granada, Spain \\
$^4$Instituto de Radioastronom\'{i}a y Astrof\'{i}sica (IRyA), UNAM Campus Morelia,
Apartado postal 3-72, 58090 Morelia, Michoac\'{a}n, Mexico\\
$^5$Instituto de Astronom\'\i a, Universidad Nacional Aut\'onoma de M\'exico,
Apdo.\ Postal 877, C.P. 22860, Ensenada, B.C., Mexico \\
}

\pubyear{2022}

\begin{document}
\label{firstpage}
\pagerange{\pageref{firstpage}--\pageref{lastpage}}
\maketitle

\begin{abstract}
We present GTC MEGARA high-dispersion integral field spectroscopic observations of the nova remnant QU\,Vul, which provide a comprehensive 3D view of this nova shell. 
The tomographic analysis of the H$\alpha$ emission reveals a complex physical structure characterized by an inhomogeneous and clumpy distribution of the material within this shell.  
The overall structure can be described as a prolate ellipsoid with an axial ratio of 1.4$\pm$0.2, a major axis inclination with the line of sight of $12^{\circ}\pm6^{\circ}$, and polar and equatorial expansion velocities $\approx$560 km~s$^{-1}$ and 400$\pm$60 km s$^{-1}$, respectively. 
The comparison of the expansion velocity on the plane of the sky with the angular expansion implies a distance of 1.43$\pm$0.23 kpc. 
The ionized mass is found to be $\approx 2\times 10^{-4}$ M$_\odot$, noting that the information on the 3D distribution of material within the nova shell has allowed us to reduce the uncertainty on its filling factor.  
The nova shell is still in its free expansion phase, which can be expected as the ejecta mass is much larger than the swept-up circumstellar medium mass.
The 3D distribution and radial velocity of material within the nova shell provide an interpretation of the so-called ``castellated'' line profiles observed in early optical spectra of nova shells, which can be attributed to knots and clumps moving radially along different directions. 
\end{abstract}




\section{Introduction}

Classical novae (CNe) outbursts occur on the surface of white dwarfs (WDs) in close binary systems. 
In these eruptive events the WD (either a carbon-oxygen or an oxygen-neon WD) accretes material from a giant or subgiant low-mass companion star via an accretion disk \citep{2008clno.book.....B}. 
When the accreted mass reaches critical conditions ($T_{\rm crit}\sim10^7$ K, $P_{\rm crit}\sim10^{20}$ dyne cm$^{-2}$), a thermonuclear runaway (TNR) occurs \citep{1986ApJ...308..721T,1998PASP..110....3G,2016PASP..128e1001S}. The outburst ejects 10$^{-5}$ to 10$^{-4}$ M$_{\odot}$ \citep{2002AIPC..637..462S} at high speeds \citep{2010AN....331..160B}. A CN event is characterized by a slow ($\sim$500–2000 km s$^{-1}$) initial wind followed by a faster (1000–4000 km s$^{-1}$) wind \citep{2008clno.book.....B}. The interaction of both winds forms a double-shock structure, with the fast wind passing through the slow wind \citep{1994MNRAS.271..155O}, producing what is known as a nova remnant. Eventually, the ejected material will expand and disperse into the interstellar medium (ISM). 

The 3D physical structure of a nova remnant results from the complex interplay of this slow-wind-fast-wind interaction, when the binary companion orbital energy and angular momentum is transferred to the fast ejecta as it proceeds through the accretion disk and circumstellar material \citep{LIVIO1990}. 
The interactions between the ejecta and the companion and accretion disk might produce intrinsically asymmetric ejecta \citep{LIVIO1990}, with larger (smaller) aspect ratio of the nova remnant for slow (fast) novae  \citep[e.g.,][]{1995MNRAS.276..353S,LLOYD1997}. Images and long-slit spectroscopic observations, interpreted by means of morpho-kinematic models, can provide 3D views of the physical structure of nova shells \citep{2000MNRAS.314..175G,2022MNRAS.512.2003S} to test these claims, although the varying conditions of observations for images and spectra, and the limited number of slit positions across the source make not always possible this investigation.

In the last years integral field spectroscopy (IFS) has opened a new window onto our understanding of the morphology, kinematics, 3D physical structure, and properties of nebular shells. 
Despite the many advantages of IFS observations, there are few works focusing on the study of nova remnants.
\cite{2009ApJ...706..738W} used multi-epoch near-IR images as well as optical Inamori Magellan Areal Camera and Spectrograph (IMACS)-IFS mounted on the 6.5m Magellan Baade telescope to discover a bipolar shell expanding around the helium nova V455\,Pup. The observations disclosed a bipolar outflow with expansion velocity of 6720$\pm$650~km~s$^{-1}$ with a pair of extremely high-velocity knots (8450$\pm$570 km~s$^{-1}$) detected at the tips of the bipolar outflows. It is also found that the nova remnant had a narrow waist at earlier times that seems to increase in size with time, thus suggesting an initial density enhancement that preceded the outburst. Its high luminosity made \cite{2009ApJ...706..738W} also suggest that the WD is massive with a high accretion rate, making it a promising supernova type Ia candidate. 
Further analysis of the kinematics and emission line spectra were presented by \cite{2014ASPC..490..115M}.
\cite{2009AJ....138.1090L} presented multi-epoch observations of the nova V723\,Cas spanning four years obtained with the near-IR IFS OSIRIS on Keck II. Interestingly, different emission lines unveiled different morphological features, with the [Si\,{\sc vi}] and [Ca\,{\sc viii}] describing an equatorial ring-like structure with a pair of polar knots, whilst the [Al\,{\sc ix}] could be attributed to a prolate spheroid structure. The authors suggest that this complex multi-structure components can be explained by independent mass ejections. \cite{2009AJ....138.1541M} used Gemini Multi-Object Spectrographs (GMOS) at the Gemini North telescope observations of HR\,Del to study its morpho-kinematic and abundances. The line maps exhibit a clumpy shell and confirm the closed hourglass morphology proposed by \cite{2003MNRAS.344.1219H}. 
Very recently \cite{2022MNRAS.511.1591T} presented an analysis of multi-instrument observations of the nova V5668\,Sgr. Multi-epoch observations obtained with Keck OSIRIS covering 3 yrs were used to study the morphology and expansion pattern of this nova shell. This nova also displays an equatorial structure with a bipolar shape. In particular, it is found that the Br$\gamma$ emission is not symmetric, whici is attributed to an accretion disk.

The studies described above reveal the great potential of IFS observations of nova shells.  
We present here high-dispersion IFS observations of QU\,Vul (a.k.a.\ Nova Vul 1984b) acquired to provide for the first time a complete view of its spatio-kinematic properties and 3D physical structure. QU\,Vul was discovered on 1984 December 22 \citep{1984IAUC.4023....1C} and confirmed afterwards spectroscopically by \cite{1987Ap&SS.130..157R}. It is a peculiar nova with an ONeMg WD \citep{1994ApJ...425..797L}, implying a unusually massive $\leq$1.0 M$_\odot$ WD \citep{2000A&A...363..647W}. Estimates of the current mass of QU\,Vul WD are in the range 0.82-0.96 M$_\odot$, which has led to the suggestion that the WD has lost $\sim$0.1 M$_\odot$ \citep{2016ApJ...816...26H}.

The nova reached a maximum visual magnitude of 5.6 mag with a decline time $t_2$ \citep[i.e., the time it takes to decline 2 mag from the peak brightness{; see}][]{1957gano.book.....G} of 20 days \citep{2010AJ....140...34S}, becoming one of the brightest fast novae. Its brightness prompted for detailed studies from early stages after its outburst, which has represented an opportunity to know the evolution of the ejected material as well as its spectroscopic properties. Its early photometric and spectral characteristics and their evolution were described by \cite{1987Ap&SS.130..157R}. In a subsequent and more detailed work, \cite{1992A&A...257..603R} analyzed the spectroscopic properties of the ejecta and confirmed that it conforms the predictions of a TNR event. 

\begin{figure}
\begin{center}
\includegraphics[width=1.0\columnwidth]{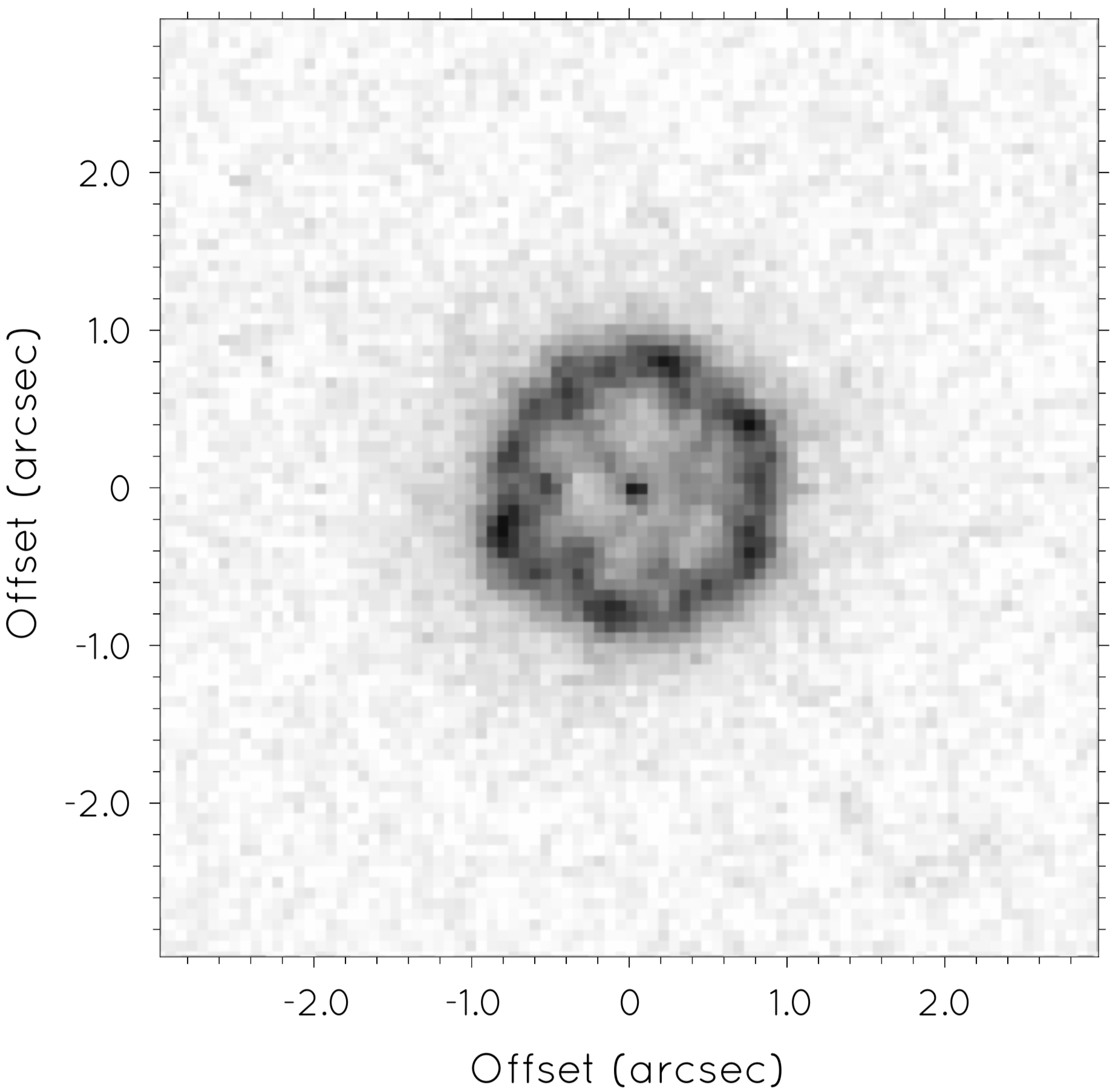}
\caption{
\textit{HST} WFPC2 F656N image of the nebular remnant around QU\,Vul obtained on 1998 January 1. 
The nebular shell is clearly resolved with an average radius $\simeq$0.8 arcsec.
A faint halo with a radius $\simeq$1.4 arcsec can be hinted in this image. 
North is up and East to the left.
}
\label{fig:hst}
\end{center}
\end{figure}

According to \cite{1992A&A...257..603R}, the nova entered the nebular phase early in April 1985 ($t-t_0 \approx 0.25$ yr) and just a few months later its radio continuum emission started showing extended emission \citep[$t-t_0 = 0.79$ and 1.36 yr;][]{1987A&A...183...38T}.  
Extended emission in the optical H$\alpha$ line was detected a few years later \citep[$t-t_0 = 9.54$ yr;][]{VAL1997}, but the first indisputable images of a resolved nebular shell were provided by the {\it Hubble Space Telescope (HST)} WFPC2 (see Fig.~\ref{fig:hst}) and NICMOS images (Prop.\ ID 7386, PI F.\ Ringwald) discussed by \citet{2000AJ....120.2007D}, reporting an ellipsoidal shell with a size of 1.7$\times$1.6 arcsec in H$\alpha$ and 1.6$\times$1.3 arcsec at 2.2 $\mu$m\footnote{
Their angular radius measurement of 5.6 arcsec based on the subtraction of the stellar PSF to the nebular one on ground-based images most likely probes the wings of the nebular PSF, not the nebular radius.}. The same {\it HST} NICMOS data were analized by \cite{2002AJ....124.2888K}, who described the nova as a nearly spherical shell with a radius of 1.07 arcsec. By 2020 the nova remnant had expanded up to a radius of 2.1 arcsec \citep{2022MNRAS.512.2003S}. 

As for the expansion velocity along the line of sight, \citet{1992A&A...257..603R} and \citet{VAL1997} reported averaged values for the full-width at half-maximum (FWHM) of a number of lines of 1380 km~s$^{-1}$ and 1190 km~s$^{-1}$, respectively, whereas \cite{2022MNRAS.512.2003S} measured an expansion velocity of 660 km~s$^{-1}$ from the resolved H$\alpha$ emission line profile.   
These expansion velocities, in conjunction with the estimates of the angular size, have been used to assess the angular expansion rate and then derive distances of 
3.6 kpc \citep{1987A&A...183...38T}, 
2.6$\pm$0.2 kpc \citep{VAL1997}, 
1.7$\pm$0.2 kpc \citep{2000AJ....120.2007D},  
3.14 kpc \citep{2002AJ....124.2888K} and 
1.4 kpc \citep{2022MNRAS.512.2003S}.  
In contrast, the distance estimate of 0.90$^{+0.35}_{-0.20}$ kpc \citep{2021AJ....161..147B} based on \textit{Gaia} Early Data Release 3 (EDR3) is smaller than all these expansion distances. 

\begin{figure*}
\begin{center}
\includegraphics[width=1.0\linewidth]{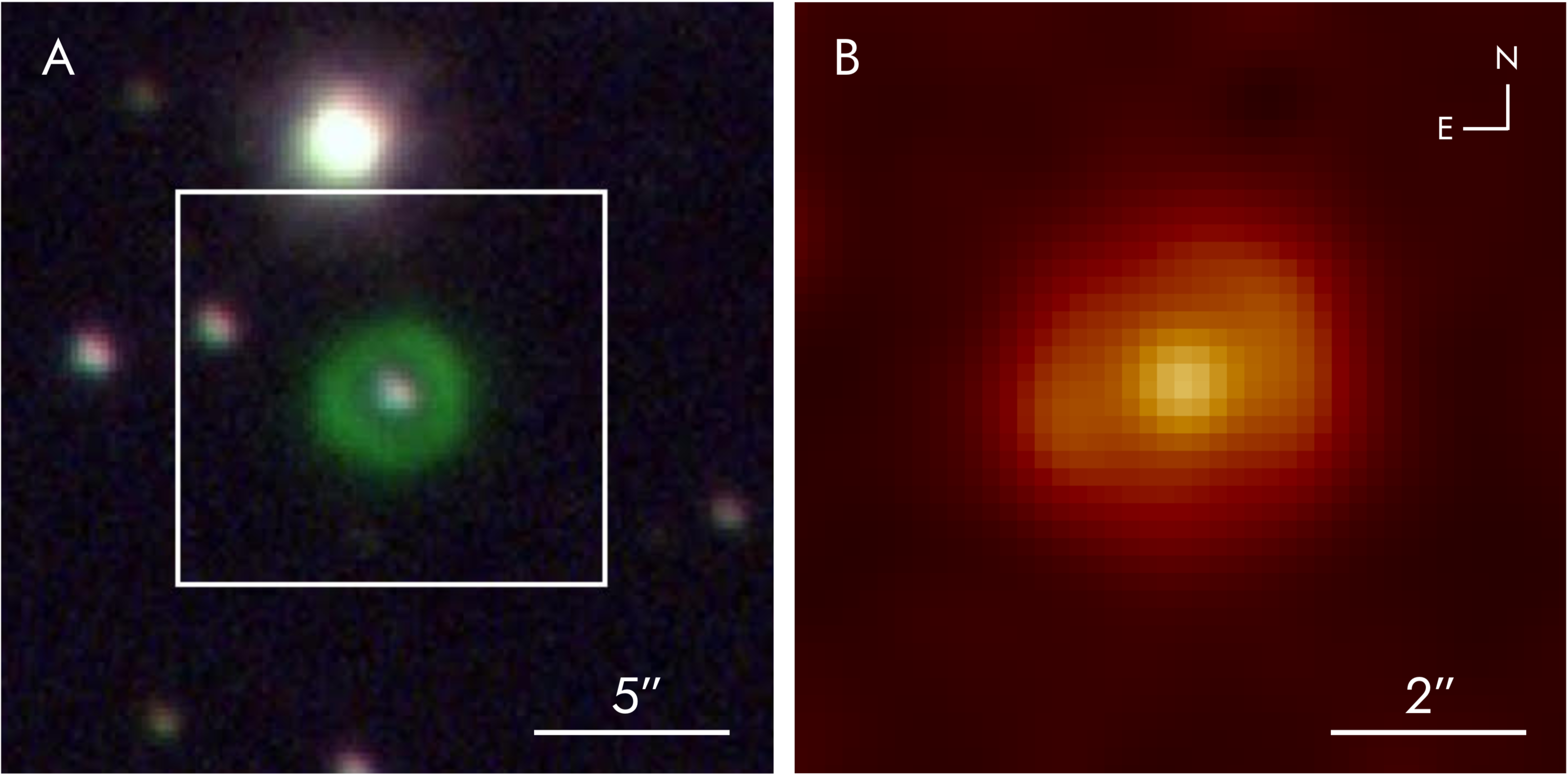}
\caption{
(A) 
NOT ALFOSC RGB composite picture of the nova QU\,Vul obtained with $g'$ SDSS $\lambda$4800 (blue) and $r'$ SDSS $\lambda$6180 (red) broad-band images, and an H$\alpha$ $\lambda$6563 narrow-band (green) image \citep[adapted from][]{2022MNRAS.512.2003S}. The area covered by the GTC MEGARA IFU (12.5$\times$11.3 arcsec) is marked by a white rectangle. (B) GTC MEGARA continuum-subtracted H$\alpha$ image of the nebular emission from QU\,Vul.
}
\label{fig:nov}
\end{center}
\end{figure*}

The discrepant expansion velocities and parallax distances\footnote{An even the differing angular sizes derived using the same {\it HST} NICMOS data by \citet{2000AJ....120.2007D} and \citet{2002AJ....124.2888K}, which illustrates the problems arising in the selection of the isophotal extent of a nebular shell by different authors.} emphasize the need for a complete study of the spatio-kinematic properties and 3D structure of the nova remnant QU\,Vul. This has been obtained with the new integral field spectroscopic observations presented here. This work is organized as follows: the observations are presented in Section \ref{2}, the results of the data analysis are described in Section \ref{3}, and a discussion is given in Section \ref{4}. A final summary is presented in Section \ref{5}.

\section{Integral Field Spectroscopic Observations}\label{2}

IFS observations of QU\,Vul were obtained on 2021 August 28 with the Multi-Espectr\'ografo en GTC de Alta Resoluci\'on para Astronom\'\i a \cite[MEGARA;][]{2018SPIE10702E..17G} attached to the 10.4 m Gran Telescopio Canarias (GTC) at the Roque de los Muchachos Observatory (ORM). We used the Large Compact Bundle (LCB) Integral Field Unit (IFU) mode which provides a FoV of 12.5$\times$11.3 arcsec$^2$ with a 567 hexagonal spaxels\footnote{Single elements of IFUs are referred to as “spatial pixels” (spaxels), the concept is used to differentiate an IFU spatial component from a detector pixel.} of maximal diameter of 0.62 arcsec. The location of the IFU field of view is shown in the left panel of Figure~\ref{fig:nov}.  
The observations were carried out with the High-resolution Volume-Phased Holographic (VPH) grism VPH665-HR (centered at 6606 {\AA}), providing a spectral range from $\sim$6405.6--6797.1 {\AA} and 0.09 {\AA} pix$^{-1}$ with a resolving power of the grating of $R=18,700$ (i.e., $\simeq$16 km s$^{-1}$). This spectral resolution is suitable to investigate the kinematics of nova shells, with typical expansion velocities $\gtrsim$500 km~s$^{-1}$ \citep{2010AN....331..160B}. Three exposures of 600 s were obtained with a seeing of 0.8 arcsec during the observations. All science frames were observed at airmasses of 1.15.

The MEGARA raw data reduction was carried out following the Data Reduction Cookbook \cite[Universidad Complutense de Madrid,][released version on 2019 July 24]{2019hsax.conf..227P}. This pipeline applies sky and bias subtraction, flat-field correction, wavelength calibration, spectra tracing and extraction. The sky subtraction is done using 56 ancillary fibres located $\approx$2.0 arcmin from the center of the IFU. The flux calibration was performed using observations of the spectro-photometric standard HR 7950. Finally, the regularization grid task {\it megararss2cube}\footnote{Tool developed by J.Zaragoza-Cardiel available at \url{https://github.com/javierzaragoza/megararss2cube}.} was used to produce a final 52$\times$58$\times$4300 data cube with 0.215~arcsec$^{2}$ spaxels.

\section{Data Analysis and Results}\label{3}

\begin{figure}
\begin{center}
\includegraphics[width=1.0\columnwidth]{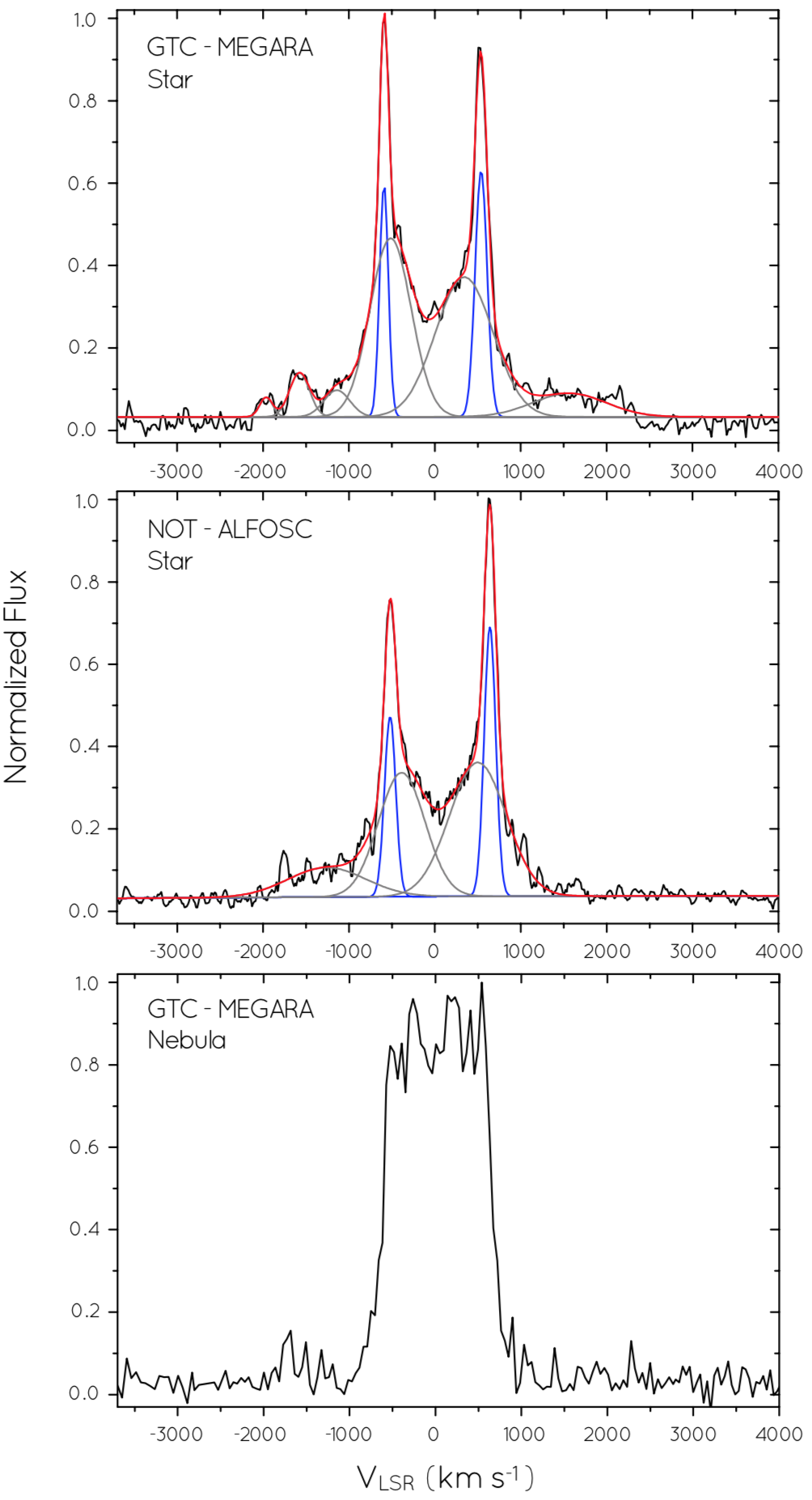}
\caption{
H$\alpha$ emission line profiles of QU\,Vul detected by GTC MEGARA (top) and NOT ALFOSC (middle) at the position of the central star, and GTC MEGARA nebular emission (bottom). The black line corresponds to the observed profile. The blue and grey lines in the stellar line profiles (top and middle) correspond to multi-Gaussian fits of the narrow nebular and broad stellar components, respectively, whereas the red line represents the total fit to the observed profile. The GTC MEGARA nebular spectrum (bottom) has a flat, ``castellated'' top. 
}
\label{fig:spec}
\end{center}
\end{figure}

The GTC MEGARA IFS observations of QU\,Vul only detect the H$\alpha$ emission line, whose spatial distribution is shown in the continuum-subtracted H$\alpha$ image presented in the right panel of Figure~\ref{fig:nov}. The H$\alpha$ line profile extracted from a circular aperture 0.5 arcsec in radius encompassing the central star of QU\,Vul is presented in the top-panel of Figure~\ref{fig:spec}.  
For comparison, the H$\alpha$ line profile extracted from NOT ALFOSC long-slit spectroscopic observations \citep[see][for a description of the NOT ALFOSC spectroscopic observations]{2022MNRAS.512.2003S} at the location of the star is also presented in the middle panel of the same figure. The stellar H$\alpha$ line profile presents two narrow peaks corresponding to the nebular emission and broad, up to $\approx$2,000 km~s$^{-1}$ wings that can be attributed to the accretion disk. Indeed the GTC MEGARA H$\alpha$ line profile of the nebular emission of QU\,Vul, extracted from an annular region with inner and outer radii 1.6 and 2.2, respectively, does not present these broad wings (Fig.~\ref{fig:spec}-bottom). A multi-Gaussian fit of the stellar H$\alpha$ emission profile has been used to derive the properties of the nebular and stellar components. A number of broad Gaussian profiles is required for the stellar component, whereas the spectral fits to the red- and blue-shifted narrow nebular components imply an expansion velocity along the line of sight $\approx$560 km~s$^{-1}$.  The possible stellar line profile variations between the NOT ALFOSC 2020 July 27 and the GTC MEGARA 2021 August 28 observations might be attributed to the orbital variation of the disk.  

\begin{figure*}
\begin{center}
\includegraphics[width=1.0\linewidth]{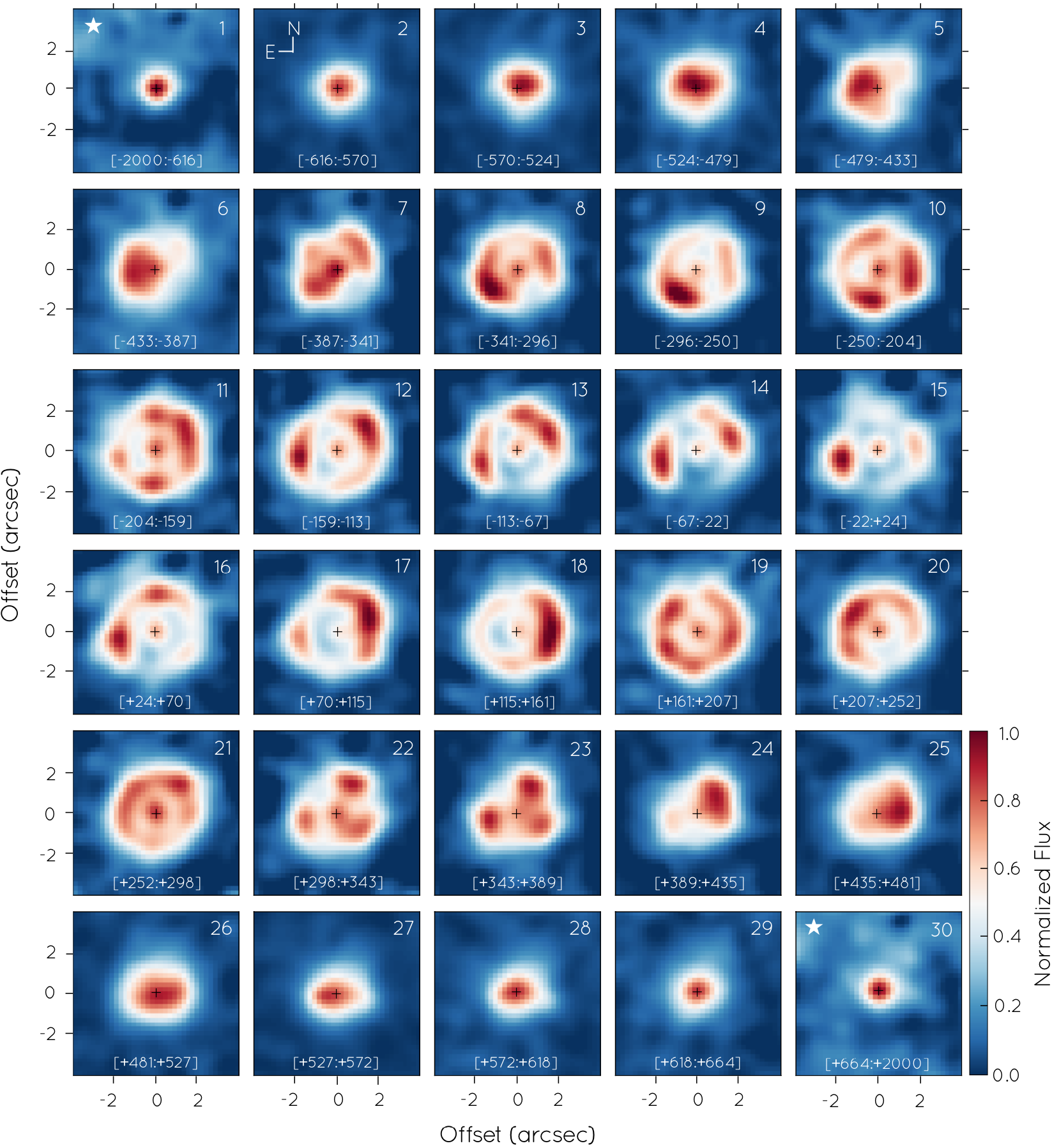}
\caption{
Normalized intensity map channels in the H$\alpha$ emission line of QU\,Vul. All channel maps but the first \#1 and last \#30 cover a velocity range of 45 km~s$^{-1}$. The velocity range of each channel map is labeled at the bottom of each map. The nebular emission spans from channel \#2 to \#29 from $-$615 to $+$665 km~s$^{-1}$. The first \#1 and last \#30 channel maps show the high-velocity wings of the H$\alpha$ emission from the central star and had been labeled by a white star. This has been used to mark the star location (black cross) in all channel maps. The line emission from the central star is also detected in channels \#7 to \#22. North is up and East to the left.
}
\label{fig:cub}
\end{center}
\end{figure*}

The H$\alpha$ line profiles of the stellar and nebular components have been used to extract the channel maps of QU\,Vul shown in Figure~\ref{fig:cub}. The nebular emission free from stellar contamination extends approximately from $-615$ to $+665$ km~s$^{-1}$. Therefore 28 channel maps with a velocity width $\approx$45 km~s$^{-1}$ have been used to map the H$\alpha$ line, from channel map \#2 to \#29 in Figure~\ref{fig:cub}. These channel maps probe different "velocity layers" of the nova shell, thus providing a tomographic view of QU\,Vul. Otherwise the first \#1 and last \#30 channel maps in Figure~\ref{fig:cub} show the location of the emission from the central star, thus providing an excellent fiducial point for comparison. 
 
Overall these channel maps describe an expanding shell, with small angular size at the velocity tips and a hollow roundish structure in the intermediate velocity channel maps \#8 to \#21. The shell is not smooth, but it rather shows a number of bright knots connected by fainter filaments. The number and spatial distribution of these knots vary notably among the different velocity channels (check, for instance, the remarkable differences between channels \#19 and \#22). The shell is neither axially symmetric, with prominent shape variations from one channel to another and with some channels particularly deviating from a circular shape (check, for instance, channels \#10, \#13, and \#15). Interestingly, the intensity peaks of the channels displaying the tips of the shell is not coincident with the position of the central star.  
The bluest tip of the shell exhibits an offset with respect to the central star $\approx$0.4 arcsec in channel \#3 towards the northwest, whereas the reddest tip in channel \#27 has a similar offset along the opposite direction. The shell can therefore be described to be prolate, with its main axis tilted along position angle (PA) $\approx$123$^\circ$. This is consistent with the nebular morphology shown in Figures~\ref{fig:hst} and \ref{fig:nov}, as well with the claim of asymmetry based on early radio observations \citep{1987A&A...183...38T}.

\begin{figure*}
\begin{center}
\includegraphics[width=1.0\linewidth]{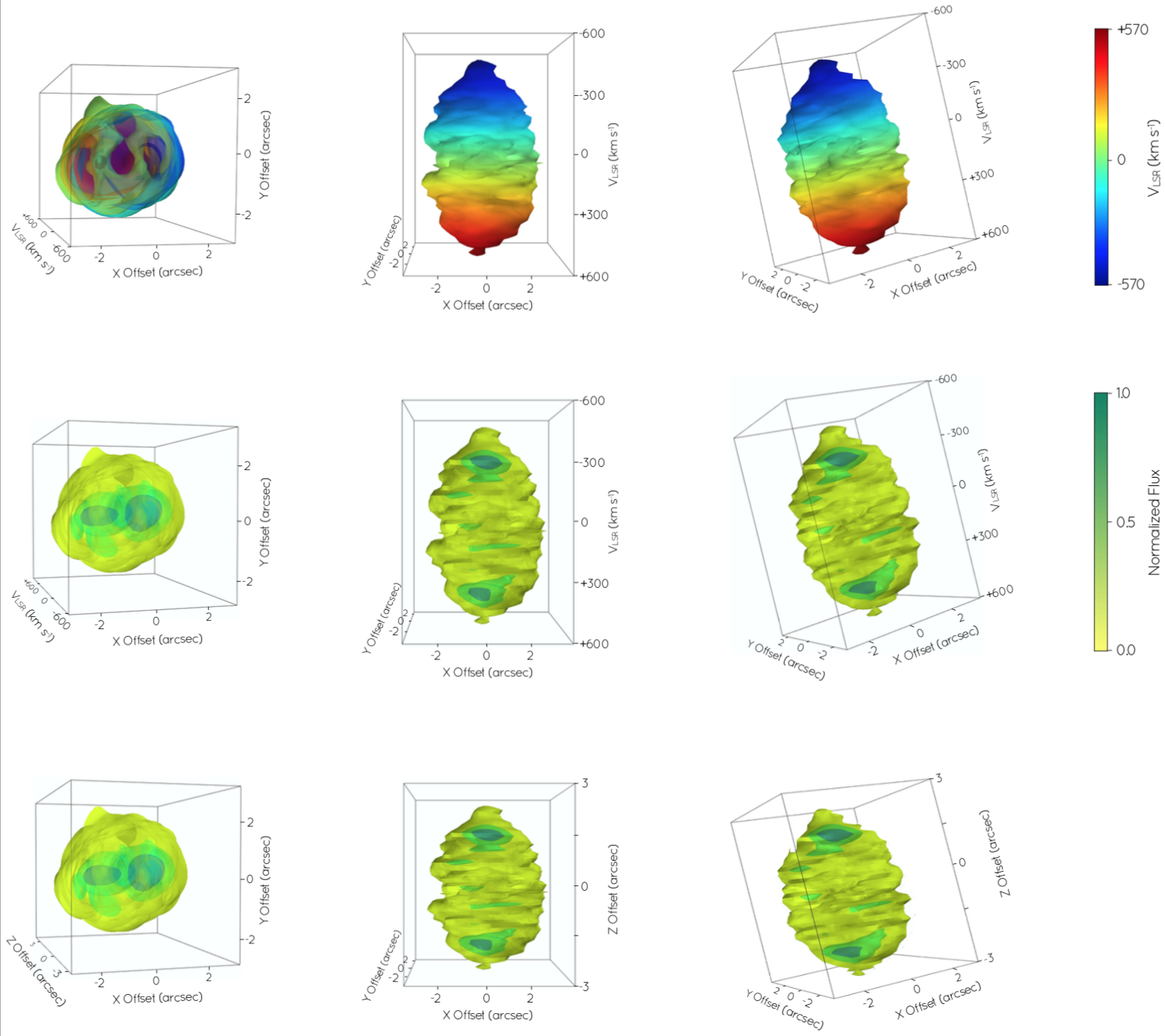}
\caption{
H$\alpha$ emission velocity-coloured (top) and intensity (middle) position-position-velocity (PPV) diagrams of QU\,Vul, and true 3D intensity distribution (bottom). The left panels show the projection along the observer's point of view (i.e., the direct image), whereas the middle and right panels show the projection from the plane of the sky along directions orthogonal and parallel to the main axis, respectively.  
The velocity range (top panels) and inhomogeneous distribution of clumps in the nebular remnant (middle and bottom panels) are appreciated in these 3D views of QU\,Vul.  
}
\label{fig:3d}
\end{center}
\end{figure*}

The complexity of the physical structure of QU\,Vul is illustrated in the position-position-velocity (PPV) diagrams shown in the top- and middle-panels of Figure~\ref{fig:3d}. The GTC MEGARA data provide a truly unique 3D view of the H$\alpha$ emission from this nova shell, which is shown from the observer point of view as the image projected on the plane of the sky (left panels), and from the plane of the sky along directions orthogonal and parallel to the main axis (central and right panels). The 3D information is displayed in full detail in the movies available in the online journal as supplementary material.

\section{Discussion}\label{4}

The 3D view of QU\,Vul presented in the previous section reveals its complexity. These data are used next to obtain key information on this nova shell. 

\subsection{3D Physical Structure}

The tomographic view of QU\,Vul shown in Figure~\ref{fig:cub} revealed it to be a prolate shell with its main axis aligned along PA $\approx$123$^\circ$. This conclusion, which is contrary to the results presented by \citet{2022MNRAS.512.2003S}, who concluded that QU\,Vul could be described as a spherical shell, emphasizes the limitations of long-slit spectroscopic observations compared to IFS observations.  

To determine the main geometrical properties of this shell, i.e., its axial ratio and inclination along the line of sight, position-velocity (PV) maps of the H$\alpha$ line have been extracted along selected PAs and compared with synthetic PVs and images of a prolate ellipsoid\footnote{Figure~\ref{fig:3d} clearly shows that the nova shell QU\,Vul is not an ellipsoid, but this simple geometrical model will be used to constrain its basic properties: aspect ratio and inclination with the line of sight.} obtained with the software {\sc shape} \citep[version 5.0;][]{2017A&C....20...87S}. Since the IFS data allows selecting any PA, the most favorable ones for this comparison  have been selected. These are the one along the projection on the plane of the sky of the main axis at PA 123$^\circ$ and that one along its orthogonal direction at PA 303$^\circ$. The comparison of the observed and synthetic H$\alpha$ line has allowed us to constrain the axial ratio of the prolate shell to 1.4$\pm$0.2 and the inclination of its main axis with the line of sight to 12$^\circ\pm$6$^\circ$. The expansion velocity at the pole is basically the velocity along the line of sight measured from the H$\alpha$ line profile, 560 km~s$^{-1}$, whereas the equatorial velocity, which is also similar to the tangential velocity on the plane of the sky, would be 400$\pm$60 km~s$^{-1}$.  

\subsection{Expansion History}

The expansion velocity on the plane of the sky could only be compared with the angular size and age of the nova to obtain its distance if a constant expansion velocity is to be assumed.  
This is indeed the case of the free-expansion phase observed for a sample of five nova shells with ages up to 130 yrs \citep{2020ApJ...892...60S}.  

There are, however, many different and not always coherent determinations of the averaged angular expansion rate of QU\,Vul: 0.080$\times$0.047 arcsec~yr$^{-1}$ along major and minor axes \citep[at age 1.36 yrs,][]{1987A&A...183...38T}, 0.099 arcsec~yr$^{-1}$ \citep[at age 9.54 yrs,][]{VAL1997}, 0.061$\times$0.057 arcsec~yr$^{-1}$ \citep[at age 14.0 yrs,][]{2000AJ....120.2007D}, and 0.059 arcsec~yr$^{-1}$ \citep[at age 35.7 yr,][]{2022MNRAS.512.2003S}. We note, however, that only the last two measurements resolved the nebular shell, whereas the previous two assumed without further discussion that the intrinsic FWHM of the emission corresponded to the shell radius. If instead the nebular radius of the unresolved shell in the two earlier measurements is equated to one half the separation of two unresolved Gaussians, it can be described in terms of the intrinsic FWHM of the source
\begin{equation}
    nebular\;radius = \frac{\rm FWHM}{2 \sqrt{\mathrm{ln}(2)}}
\end{equation}
and similar expansion rates of 0.096--0.057 arcsec~yr$^{-1}$ \citep{1987A&A...183...38T} and 0.059 arcsec~yr$^{-1}$ \citep{VAL1997} are obtained. It can thus be concluded that the angular expansion rate of QU\,Vul has remained constant at a value $\approx$0.060 arcsec~yr$^{-1}$ since its outburst in 1984.  

As for the velocity along the line of sight, earlier measurements implied expansion velocities larger than the observed 560 km~s$^{-1}$. \citet{1992A&A...257..603R} reported the appearance of a number of blue-shifted absorption components of different lines  at $-$1350, $-$850, and $-$680 km~s$^{-1}$.  
These disappeared soon in the nebular phase and can be interpreted as high-velocity clumps ejected at the time of the nova explosion. \citet{1992A&A...257..603R} also derived expansion velocities in the range 1570 to 1380 km~s$^{-1}$ for different epochs from the FWHM of a number of emission lines, but for H$\alpha$, which implied even larger expansion velocities. These results are consistent with those presented by \citet{VAL1997} with data obtained a few years later that implied expansion velocities of 1190 km~s$^{-1}$ also from the FWHM of emission line profiles.  

It must be noted that the emission line profiles presented by these authors can be described as "castellated", i.e., they have a rectangular shape with multiple peaks at the top that resemble the wall and battlements of a castle. This is exactly the same shape of the H$\alpha$ line profile of the whole nebular emission of QU\,Vul at the bottom of Figure~\ref{fig:spec}, whose FWHM, $\approx1200$ km~s$^{-1}$, is similar to the values reported by \citet{1992A&A...257..603R} and \citet{VAL1997}. This H$\alpha$ line profile would be the one observed for a spatially unresolved shell \citep[as it is the case of the earliest line profiles of novae, e.g.][]{1991IAUC.5236....1I}. This spectral shape can be interpreted in view of our spatially and spectroscopically resolved data: the "wall" and its "battlements" result from the combination of the emission from multiple knots and filaments moving at different velocities along the line of sight. 

The result described above illustrate that the FWHM of the castellated line profiles observed early in the evolution of nova shells should not be used to estimate its expansion velocity. Instead we propose that the bluest and reddest peaks of spatially unresolved castellated profiles should be used to define lower limits for the expansion velocity. The line profiles presented by \citet{1992A&A...257..603R} would then indicate that the expansion velocity of QU\,Vul in 1987 was larger than 420 km~s$^{-1}$ and larger than 505 km~s$^{-1}$ in 1990. The comparison of these values with the one measured here is thus consistent with a constant expansion velocity for QU\,Vul along the line of sight.  

It must also be remarked that the larger FWHM of the H$\alpha$ line reported by \citet{VAL1997} is most likely caused by the contamination of the high-velocity wings of the stellar/disk emission, as illustrated by the comparison of the top and bottom panels in Figure~\ref{fig:spec}. It is also worth noting that the changes in the castellated profiles described by \citet{1992A&A...257..603R} most likely correspond to the brightening and dimming of knots with time caused by shocks in the expanding shell, similar to those observed in SN\,1987A \citep{1991ApJ...369L..63J}. 

\subsection{Distance}

The free expansion at constant velocity of QU\,Vul has been found above to be the most simple interpretation of the multi-epoch imaging and spectroscopic observations obtained so far. Therefore the equation 
\begin{equation}
    \tau = 4750 ~\frac{\theta \times d}{v} 
\end{equation}
relating the age $\tau$ in years, the angular radius $\theta$ in arcsec, the expansion velocity on the plane of the sky $v$ in km~s$^{-1}$, and the distance $d$ in kpc can be used. For an age of 35.7 years, the angular radius of 2.1$\pm$0.1 arcsec derived by \citet{2022MNRAS.512.2003S} and the expansion velocity of 400$\pm$60 km~s$^{-1}$ presented here imply a distance of 1.43$\pm$0.23 kpc. 

The distance derived here is substantially smaller than those previously reported, which are affected by the inappropriate assumptions on the expansion velocity and angular sizes described above. On the other hand, this distance is closer (although still larger) to the one derived from \textit{Gaia} EDR3 of 0.90$^{+0.35}_{-0.20}$ kpc \citep{2021AJ....161..147B}, being both distances consistent at 1$\sigma$. Note that the determination of \textit{Gaia} distances for stars surrounded by unresolved diffuse emission has been questioned \citep[e.g.,][]{2018A&A...616L...2K,2020MNRAS.499.2959H}, as red targets with nebular remnant or high-brightness nebulosities show a much larger parallax dispersion, systematically underestimating the errors of objects with circumstellar material in the \textit{Gaia} DR2 database.

\subsection{Time Evolution of the H$\alpha$ Luminosity}

The H$\alpha$ luminosity of a nova shell can be expected to decline with time, as the shell expansion implies a decrease in density, while the nova material further cools down. The H$\alpha$ luminosity of QU\,Vul has been studied from early stages after its outburst in 1984 and we can compare those with the H$\alpha$ luminosity estimated in this study to investigate its time evolution. 

The first report of the ejecta luminosity \citep{1992A&A...257..603R} implied values for the intrinsic F(H$\alpha$) of $7.9\times10^{-12}$ erg cm$^{-2}$ s$^{-1}$ in August 1987, $3.4\times10^{-12}$ erg cm$^{-2}$ s$^{-1}$ in September 1988, and $1.3\times10^{-12}$ erg cm$^{-2}$ s$^{-1}$ in August 1990.  
We note that a value of 5.7$\pm$0.7 has been used for the H$\alpha$ to H$\beta$ ratio to derive the values listed above. In August 1994, the intrinsic H$\alpha$ flux had decreased to $6.8\times10^{-13}$ erg cm$^{-2}$ s$^{-1}$ \citep{VAL1997}, with an H$\alpha$ to H$\beta$ ratio of 5.5. The last measurement dates from January 1998, as obtained by \citet{2000AJ....120.2007D} using a \textit{HST} H$\alpha$ image at that time. The intrinsic H$\alpha$ flux can be worked out to be $2.0\times10^{-13}$ erg~cm$^{-2}$ s$^{-1}$. The intrinsic H$\alpha$ flux derived from our GTC MEGARA observations is  $8.4\times10^{-14}$ erg cm$^{-2}$ s$^{-1}$. Both this value and that of \citet{2000AJ....120.2007D} have been corrected from absorption using the extinction value given by \citet{2018MNRAS.476.4162O} of $E(B-V)=0.55$ or c(H$\beta$)=0.78, which implies a ratio of the observed to the intrinsic H$\alpha$ flux of 3.6. The present H$\alpha$ luminosity of QU\,Vul is thus estimated to be $2.1\times10^{31}$ erg s$^{-1}$.

\begin{figure}
\begin{center}
\includegraphics[width=1.0\columnwidth]{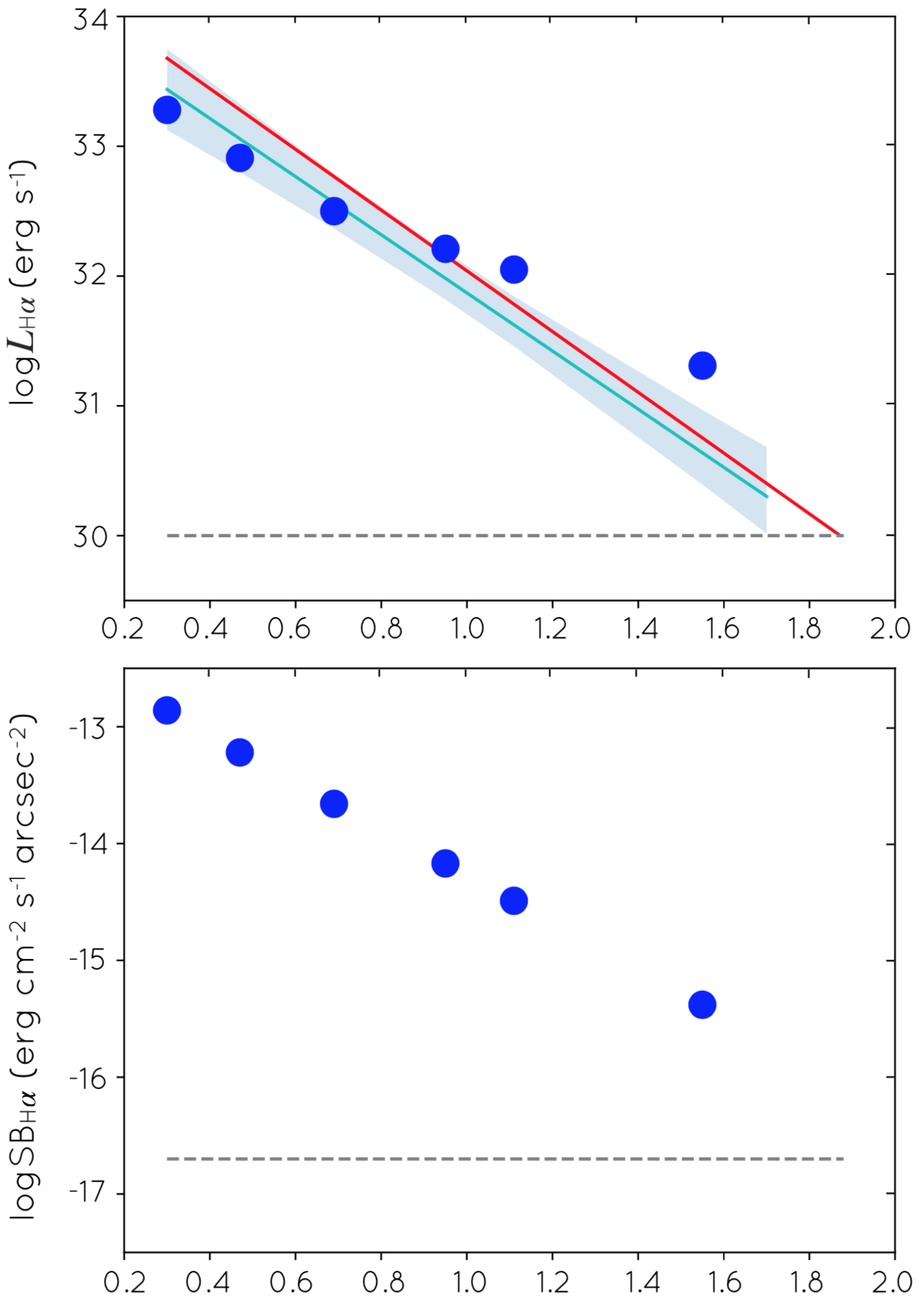}
\caption{
Time evolution of the QU\,Vul intrinsic H$\alpha$ luminosity (top) and observed H$\alpha$ surface brightness (bottom). The red and cyan lines in the top panel describe the time evolution of the intrinsic H$\alpha$ luminosity derived by \citet{2001JAD.....7....6D} and \citet{2020A&A...641A.122T}, respectively. The light-blue band in the \citet{2020A&A...641A.122T} relationship represents the uncertainty described by the authors. The dotted horizontal lines mark the detection limits of the H$\alpha$ luminosity of 10$^{30}$ erg s$^{-1}$ \citep[top,][]{2001JAD.....7....6D} and IPHAS surface brightness \citep[bottom,][]{2005MNRAS.362..753D}. 
}
\label{fig:lum}
\end{center}
\end{figure}

The time evolution of the H$\alpha$ luminosity of QU\,Vul is shown in the top panel of Figure~\ref{fig:lum}. The H$\alpha$ luminosity declines quickly at early times and much more slowly at late times. This behaviour can be compared to that revealed by (the very few) studies of the H$\alpha$ luminosity evolution of nova shells. \cite{2001JAD.....7....6D} carried out a detailed study on the behavior of the  H$\alpha$, H$\beta$, and [O~{\sc iii}] $\lambda$5007 luminosities of nova remnants over time collecting information for a sample of novae of different speed classes. For fast novae, as it is the case of QU\,Vul, the H$\alpha$ luminosity is derived to vary as: 
\begin{equation}
    \log L_{{\rm H}\alpha} = (34.38\pm0.10) - (2.34\pm0.11) \log{(t-t_0)}.
\end{equation}
A more recent investigation of the long-term evolution of the H$\alpha$ and [O~{\sc iii}] luminosities of nova shells in comparison with their light curve and nova speed class was presented by \citet{2020A&A...641A.122T}. They propose that the H$\alpha$ luminosity as a function of time can be described by different stages, where it is likely that the evolution of luminosity in the first phases is dominated by shell expansion and changes in volume and density, while in older nova shells the interaction with the ISM takes an important role. Their relationship for the time evolution of the H$\alpha$ luminosity of fast novae is: 
\begin{equation}
    \log L_{{\rm H}\alpha} = (34.11\pm0.23) - (2.24\pm0.24) \log{(t-t_0)}, 
\end{equation}

which is noted to have a significant scatter ($\sigma$=0.69). The comparison of the time evolution of the H$\alpha$ luminosity of QU\,Vul with these relationships in Figure~\ref{fig:lum}-top shows an excellent agreement. Our measurement of its recent H$\alpha$ luminosity indicates that it might have entered a phase of slower evolution, although the slope of the decay observed in QU\,Vul is still consistent with the trends reported by \citet{2001JAD.....7....6D} and \citet{2020A&A...641A.122T} within the uncertainty of these relationships. 

The long-term decline of the H$\alpha$ luminosity of QU\,Vul is obvious in Figure~\ref{fig:lum}-top.  
The decline of the H$\alpha$ surface brightness is even deeper in the bottom panel of the same figure (three orders of magnitude in 35 years), as the nova shell expands with time. This situation raises the question of how long a nova shell would be detectable after the eruption before dissipating in the ISM. After accounting for distances, type of nova, reddening, remnant luminosity and telescope resolution, \cite{2001JAD.....7....6D} concluded that the lowest H$\alpha$ luminosity in their sample was 10$^{30}$ erg s$^{-1}$. Meanwhile the detection limit of IPHAS, the INT Photometric H$\alpha$ Survey \citep{2005MNRAS.362..753D}, is $2\times10^{-17}$ erg~cm$^{-2}$~s$^{-1}$~arcsec$^{-2}$.  
These detection limits are plotted on Figure~\ref{fig:lum}, where it can be foreseen that QU\,Vul would faint below these at an age $\approx$80 years, with the flattening of the emission evolution in the last years suggesting a longer visibility time period up to an age of $\approx$200 years. These are typical ages of nova shells \citep{2015MNRAS.451.2863S}. 

\subsection{Ionized Mass and Kinetic Energy}

The ionized mass of a nebula can be determined by estimating its average electron density ($N_{\rm e}$) from the intrinsic H$\alpha$ flux ($F_{{\rm H}\alpha}$) using, for instance, the following expression adapted from \citet{1970Ap&SS...6..183M}: 
\begin{equation}
    N_{\rm e} = 1.2 \sqrt{\frac{4 \pi d^2 F_{\rm H\alpha}}{\varepsilon_{\rm H\alpha} V f}}
\end{equation}
where $d$ is the distance to the nebula, $V$ its volume, $\varepsilon_{\rm H\alpha}$ is the emission coefficient, whose value is $4\times 10^{-25}$ erg cm$^3$ s$^{-1}$ for a plasma at an electron temperature $T_{\rm e}$ of 10,000 K \citep{1968IAUS...34..162B}, $N_{\rm e}$/$N_{\rm p}$ is assumed to be 1.2, and $f$ is the filling factor. The latter can be described as $f=a \times b$, where the first term \textit{a} is the fraction of the volume of the shell given by its thickness (i.e., the macroscopic component of the filling factor) and the second term \textit{b} represents the fraction of the volume covered by condensations (i.e., the microscopic component of the filling factor), which are both smaller than unity. The average electron density of QU\,Vul would be $370 \times f^{-1/2}$ cm$^{-3}$.

The total ionized mass of the shell is then computed as: 
\begin{equation}
M_{\rm shell} = \mu \, m_{\rm p} \, N_{\rm p} \, V \, f = \mu \, m_{\rm p} \, \sqrt{\frac{4 \pi d^2 F_{\rm H\alpha} V f}{\varepsilon_{\rm H\alpha}}}
\label{eq.mass}
\end{equation}
where $\mu$ is the mean molecular weight, which can be assumed to be 1.44 for a He/H solar ratio.  
Eq.~\ref{eq.mass} is typically evaluated for the whole shell using its total volume and H$\alpha$ flux, which would imply an ionized mass $\approx 2.4\times10^{-4} f^{1/2}~M_{\odot}$. Very interestingly, the GTC MEGARA IFS observations of QU\,Vul allow us the possibility to actually evaluate it for each volume element of the 3D data cube. This includes information on the true 3D structure and the inhomogeneous clumpy distribution of the ejected material in the nebular shell (as illustrated in the channel maps shown in Fig.~\ref{fig:cub}), reducing the uncertainty in the macroscopic term $a$ of the filling factor, which is effectively determined. The volume element of the data cube is define by the size of the spatial pixel ($0.215\times0.215$ arcsec) and the wavelength range, which was rebinned to three pixels along the spectral direction (0.27 \AA\ $\approx$ 12 km~s$^{-1}$) to increase the SNR. The latter corresponds to an angular size of 0.065 arcsec, for an ellipsoid with an axial ratio of 1.4, an equatorial radius of 2.1 arcsec, and a polar expansion velocity of 560 km~s$^{-1}$. Accordingly, each data cube element has a volume of $3.0\times10^{46}$ cm$^3$.  

The H$\alpha$ flux of each volume element was corrected from extinction using the factor 3.6 corresponding to a c(H$\beta$) extinction coefficient of 0.78 and the distribution of the H$\alpha$ intrinsic intensities and thus of the ionized mass in each volume element computed. The total ionized mass is then obtained by adding the mass in each volume element, which results in a value of $1.9\times10^{-4} b^{1/2}~M_{\odot}$, similar to that reported in previous works \citep{1987A&A...183...38T,1992ApJ...398..651S}. We emphasize that the ionized mass estimate presented here does not require to assume a value of the macroscopic component of the filling factor $a$ as the H$\alpha$ flux is directly measured for every volume element of the data cube. Indeed, a comparison of this mass and the average mass computed above implies a value of 0.8 for $a$.  

The ionized mass reported here can be compared with the typical mass ejected in nova outbursts of $\sim10^{-4}~M_{\odot}$ \citep{2002AIPC..637..462S} for typical values of the microscopic component of the filling factor $b$ in the range 0.1 to unity. The total volume reached by the nova shell implies a swept-up mass of $2.5\times10^{-7}~M_{\odot}$ for an assumed value of 0.55 cm$^{-3}$ for the density of the circumstellar medium \citep{2016A&A...594A.116H}. The mass of the nova ejecta is indeed much greater than the mass of the swept ISM, by a factor $760\times b^{1/2}$, which is consistent with its free expansion phase. The latter is also supported by the kinetic energy of QU\,Vul, $E_{\rm kin}=3.1\times10^{44}$ erg, which has been derived by adopting an expansion velocity weighted between the radial velocity and the velocity on the plane of the sky. The kinetic energy of nova shells ranges from 10$^{43}$ to 10$^{45}$ erg for slow and fast novae, respectively \citep{1978ARA&A..16..171G}. 

\section{Summary}\label{5}

Integral field spectroscopic observations of nova shells have the potential to determine their basic properties and to unveil fine details of their structure, yet very few studies of this type have been carried out so far. Here we present a detailed morpho-kinematic analysis of the nova shell QU\,Vul (a.k.a., Nova Vul 1984b) using IFS observations obtained with MEGARA at the GTC. 

The observations detect emission only from the H$\alpha$ line, which has been used to obtain a 3D view of the nebula in the position-position-velocity space. This tomographic view reveals that QU\,Vul can basically be described as a tilted prolate shell with an  inhomogeneous and clumpy structure. A spatio-kinematic model to these data, in conjunction with available intermediate-resolution NOT ALFOSC spectra, consisting of a prolate ellipsoid with homologous expansion implies polar and equatorial semiaxes of 2.9 and 2.1 arcsec, respectively, for an axial ratio of 1.4$\pm$0.2, an inclination of its main axis with the line of sight at $12^{\circ}\pm6^{\circ}$, and polar and equatorial velocities $\approx$560 km~s$^{-1}$ and 400$\pm$60 km s$^{-1}$, respectively. The position-position-velocity data cube provided by the IFS observations have then been used to obtain the true 3D physical structure of QU\,Vul. 

After checking that the expansion rate of QU\,Vul since the nova event in December 1984 has been constant at a rate $\approx$0.060 arcsec~yr$^{-1}$, i.e., the nova ejecta is still in its free expansion phase, the tangential velocity in the plane of the sky of 400$\pm$60 km s$^{-1}$ has been used to derive a distance of 1.43$\pm$0.23 kpc. In checking the possible variation in time of the expansion rate, we note the many discrepant early results on the nebular size and its expansion velocity along the line of sight.  
The former arise from measurements of angular size from images that do not resolve the nebular shell properly, the latter from the assumption that the FWHM of ``castellated'' line profiles can be used to derive the expansion velocity. It is proposed that one half of the velocity separation between the bluest and reddest peaks of line profiles from spatially unresolved nova shells provide a lower limit for its expansion velocity.  

The H$\alpha$ flux measured in the IFS observations has been used to determine an ionized mass for QU\,Vul of $1.9\times10^{-4} M_\odot$. Since this mass is obtained from a 3D data cube, it already accounts for the macroscopic component of the filling factor, for which a value of 0.8 is derived. It is noteworthy to remark that the mass of the ejected material is several hundreds much larger than the mass of the swept-up ISM, which is consistent with a free expansion.

To sum up, this work corroborates the expected potential of IFS observations of nova shells. The overall 3D structure (axial ratio and inclination) and fine details (clumps distribution and shell thickness) of the nova shell of QU\,Vul have been clearly established and revealed. The 3D structure allows a correct interpretation of the expansion velocity on the plane of the sky and determination of the distance.  
It also allows interpreting the emission line profiles of unresolved nova shells, which are found to be dominated by the emission from bright clumps moving at different velocities along the line of sight.  
Finally the 3D information on the distribution of the H$\alpha$ emission within the nebular shell allows computing the ionized mass without any assumption on the value of the macroscopic filling factor.  

\section*{Acknowledgments} 

E.S.\ acknowledges support from Universidad de Guadalajara and Consejo Nacional de Ciencia y Tecnolog\'{i}a (CONACyT) for a student scholarship. M.A.G.\ acknowledges support of grant PGC 2018-102184-B-I00 of the Ministerio de Educación, Innovación y Universidades cofunded with FEDER funds. G.R.-L.\ acknowledges support from Universidad de Guadalajara, CONACyT grant 263373 and Programa para el Desarrollo Profesional Docente (PRODEP, Mexico). J.A.T.\ acknowledges funding from the Marcos Moshinsky Foundation (Mexico) and Dirección General de Asuntos del PersonalAcadémico (DGAPA), Universidad Nacional Autónoma de México, through grants Programa de Apoyo a Proyectos de Investigación e Innovación Tecnológica (PAPIIT) IA101622. L.S.\ acknowledges support from PAPIIT grant IN110122. The data presented here were obtained in part with ALFOSC, and provided by the Instituto de Astrofísica de Andalucía (IAA) uder a joint agreement with the University of Copenhagen and NOTSA. The GTC Science Operations team is acknowledged for scheduling the GTC MEGARA observations under the stringent conditions demanded by this program.

The authors are indebted to Drs.\ Sara Cazzoli and Alessandro Ederoclite for their careful reading of the manuscript and their valuable comments and suggestions. 

\section*{Data Availability} 

The data underlying this work are available in the article. The data files will be shared on request to the first author.





\begin{thebibliography}{99}
  
\bibitem[Bailer-Jones et al.(2021)]{2021AJ....161..147B} Bailer-Jones, C.~A.~L., Rybizki, J., Fouesneau, M., et al.\ 2021, \aj, 161, 147. doi:10.3847/1538-3881/abd806

\bibitem[Bode \& Evans(2008)]{2008clno.book.....B} Bode, M.~F. \& Evans, A.\ 2008, Classical Novae, 2nd Edition. Edited by M.F. Bode and A. Evans. Cambridge Astrophysics Series, No. 43, Cambridge: Cambridge University Press, 2008.

\bibitem[Bode(2010)]{2010AN....331..160B} Bode, M.~F.\ 2010, Astronomische Nachrichten, 331, 160. doi:10.1002/asna.200911319

\bibitem[Boyarchuk et al.(1968)]{1968IAUS...34..162B} Boyarchuk, A.~A., Gershberg, R.~E., Godovnikov, N.~V., et al.\ 1968, Planetary Nebulae, 34, 162

\bibitem[Collins et al.(1984)]{1984IAUC.4023....1C} Collins, P., Hurst, G.~M., Wils, P., et al.\ 1984, \iaucirc, 4023

\bibitem[della Valle et al.(1997)]{VAL1997}
della Valle, M., Gilmozzi, R., Bianchini, A., \& Esenoglu, H.\ 1997,
\aap, 325, 1151 
  
\bibitem[Downes \& Duerbeck(2000)]{2000AJ....120.2007D} Downes, R.~A. \& Duerbeck, H.~W.\ 2000, \aj, 120, 2007. doi:10.1086/301551

\bibitem[Downes et al.(2001)]{2001JAD.....7....6D} Downes, R.~A., Duerbeck, H.~W., \& Delahodde, C.~E.\ 2001, Journal of Astronomical Data, 7, 6

\bibitem[Drew et al.(2005)]{2005MNRAS.362..753D} Drew, J.~E., Greimel, R., Irwin, M.~J., et al.\ 2005, \mnras, 362, 753. doi:10.1111/j.1365-2966.2005.09330.x

\bibitem[Gallagher \& Starrfield(1978)]{1978ARA&A..16..171G} Gallagher, J.~S. \& Starrfield, S.\ 1978, \araa, 16, 171. doi:10.1146/annurev.aa.16.090178.001131

\bibitem[Gaposchkin(1957)]{1957gano.book.....G} Gaposchkin, C.~H.~P.\ 1957, Amsterdam, North-Holland Pub. Co.; New York, Interscience Publishers, 1957.

\bibitem[Gehrz et al.(1998)]{1998PASP..110....3G} Gehrz, R.~D., Truran, J.~W., Williams, R.~E., et al.\ 1998, \pasp, 110, 3. doi:10.1086/316107

\bibitem[Gil de Paz et al.(2018)]{2018SPIE10702E..17G} Gil de Paz, A., Carrasco, E., Gallego, J., et al.\ 2018, \procspie, 10702, 1070217. doi:10.1117/12.2313299

\bibitem[Gill \& O'Brien(2000)]{2000MNRAS.314..175G} Gill, C.~D. \& O'Brien, T.~J.\ 2000, \mnras, 314, 175. doi:10.1046/j.1365-8711.2000.03342.x

\bibitem[Hachisu \& Kato(2016)]{2016ApJ...816...26H} Hachisu, I. \& Kato, M.\ 2016, \apj, 816, 26. doi:10.3847/0004-637X/816/1/26

\bibitem[Harman \& O'Brien(2003)]{2003MNRAS.344.1219H} Harman, D.~J. \& O'Brien, T.~J.\ 2003, \mnras, 344, 1219. doi:10.1046/j.1365-8711.2003.06906.x

\bibitem[Harvey et al.(2020)]{2020MNRAS.499.2959H} Harvey, E.~J., Redman, M.~P., Boumis, P., et al.\ 2020, \mnras, 499, 2959. doi:10.1093/mnras/staa2896

\bibitem[HI4PI Collaboration et al.(2016)]{2016A&A...594A.116H} HI4PI Collaboration, Ben Bekhti, N., Flöer, L., et al.\ 2016, \aap, 594, A116. doi:10.1051/0004-6361/201629178

\bibitem[\protect\citeauthoryear{Iijima et al.}{1991}]{1991IAUC.5236....1I} 
Iijima T., Gehrz R.~D., Jones T.~J., Lawrence G., Sivaraman K.~R., Prabhu T.~P., Ghosh K.~K., et al., 1991, IAUC, 5236

\bibitem[Jakobsen et al.(1991)]{1991ApJ...369L..63J} Jakobsen, P., Albrecht, R., Barbieri, C., et al.\ 1991, \apjl, 369, L63. doi:10.1086/185959

\bibitem[Kimeswenger \& Barr{\'\i}a(2018)]{2018A&A...616L...2K} Kimeswenger, S. \& Barr{\'\i}a, D.\ 2018, \aap, 616, L2. doi:10.1051/0004-6361/201833647

\bibitem[Krautter et al.(2002)]{2002AJ....124.2888K} Krautter, J., Woodward, C.~E., Schuster, M.~T., et al.\ 2002, \aj, 124, 2888. doi:10.1086/343833

\bibitem[Livio et al.(1990)]{LIVIO1990}
Livio, M., Shankar, A., Burkert, A., et al.\ 1990,
\apj, 356, 250. doi:10.1086/168836

\bibitem[Livio \& Truran(1994)]{1994ApJ...425..797L} Livio, M. \& Truran, J.~W.\ 1994, \apj, 425, 797. doi:10.1086/174024
  
\bibitem[Lloyd et al.(1997)]{LLOYD1997}
Lloyd, H.~M., O'Brien, T.~J., \& Bode, M.~F.\ 1997,
\mnras, 284, 137. doi:10.1093/mnras/284.1.137
  
\bibitem[Lyke \& Campbell(2009)]{2009AJ....138.1090L} Lyke, J.~E. \& Campbell, R.~D.\ 2009, \aj, 138, 1090. doi:10.1088/0004-6256/138/4/1090

\bibitem[Macfarlane et al.(2014)]{2014ASPC..490..115M} Macfarlane, S., Steeghs, D., \& Woudt, P.\ 2014, Stellar Novae: Past and Future Decades, 490, 115

\bibitem[Moraes \& Diaz(2009)]{2009AJ....138.1541M} Moraes, M. \& Diaz, M.\ 2009, \aj, 138, 1541. doi:10.1088/0004-6256/138/6/1541

\bibitem[Mustel \& Boyarchuk(1970)]{1970Ap&SS...6..183M} Mustel, E.~R. \& Boyarchuk, A.~A.\ 1970, \apss, 6, 183. doi:10.1007/BF00651221

\bibitem[O'Brien et al.(1994)]{1994MNRAS.271..155O} O'Brien, T.~J., Lloyd, H.~M., \& Bode, M.~F.\ 1994, \mnras, 271, 155. doi:10.1093/mnras/271.1.155

\bibitem[{\"O}zd{\"o}nmez et al.(2018)]{2018MNRAS.476.4162O} {\"O}zd{\"o}nmez, A., Ege, E., G{\"u}ver, T., et al.\ 2018, \mnras, 476, 4162. doi:10.1093/mnras/sty432

\bibitem[Pascual et al.(2019)]{2019hsax.conf..227P} Pascual, S., Cardiel, N., Gil de Paz, A., et al.\ 2019, Highlights on Spanish Astrophysics X, 227

\bibitem[Rosino \& Iijima(1987)]{1987Ap&SS.130..157R} Rosino, L. \& Iijima, T.\ 1987, \apss, 130, 157. doi:10.1007/BF00654989

\bibitem[Rosino et al.(1992)]{1992A&A...257..603R} Rosino, L., Iijima, T., Benetti, S., et al.\ 1992, \aap, 257, 603

\bibitem[Sahman et al.(2015)]{2015MNRAS.451.2863S} Sahman, D.~I., Dhillon, V.~S., Knigge, C., et al.\ 2015, \mnras, 451, 2863. doi:10.1093/mnras/stv1150

\bibitem[Saizar et al.(1992)]{1992ApJ...398..651S} Saizar, P., Starrfield, S., Ferland, G.~J., et al.\ 1992, \apj, 398, 651. doi:10.1086/171890

\bibitem[Santamar{\'\i}a et al.(2020)]{2020ApJ...892...60S} Santamar{\'\i}a, E., Guerrero, M.~A., Ramos-Larios, G., et al.\ 2020, \apj, 892, 60. doi:10.3847/1538-4357/ab76c5

\bibitem[Santamar{\'\i}a et al.(2022)]{2022MNRAS.512.2003S} Santamar{\'\i}a, E., Guerrero, M.~A., Zavala, S., et al.\ 2022, \mnras, 512, 2003. doi:10.1093/mnras/stac563

\bibitem[Shafter(2002)]{2002AIPC..637..462S} Shafter, A.~W.\ 2002, Classical Nova Explosions, 637, 462. doi:10.1063/1.1518246
  
\bibitem[Slavin et al.(1995)]{1995MNRAS.276..353S}
  Slavin, A.~J., O'Brien, T.~J., \& Dunlop, J.~S.\ 1995,
  \mnras, 276, 353. doi:10.1093/mnras/276.2.353

\bibitem[Strope et al.(2010)]{2010AJ....140...34S} Strope, R.~J., Schaefer, B.~E., \& Henden, A.~A.\ 2010, \aj, 140, 34. doi:10.1088/0004-6256/140/1/34
  
\bibitem[Starrfield et al.(2016)]{2016PASP..128e1001S} Starrfield, S., Iliadis, C., \& Hix, W.~R.\ 2016, \pasp, 128, 51001. doi:10.1088/1538-3873/128/963/051001

\bibitem[Steffen \& Koning(2017)]{2017A&C....20...87S} Steffen, W. \& Koning, N.\ 2017, Astronomy and Computing, 20, 87. doi:10.1016/j.ascom.2017.06.002

\bibitem[Takeda et al.(2022)]{2022MNRAS.511.1591T} Takeda, L., Diaz, M., Campbell, R.~D., et al.\ 2022, \mnras, 511, 1591. doi:10.1093/mnras/stac097

\bibitem[Tappert et al.(2020)]{2020A&A...641A.122T} Tappert, C., Vogt, N., Ederoclite, A., et al.\ 2020, \aap, 641, A122. doi:10.1051/0004-6361/202037913

\bibitem[Taylor et al.(1987)]{1987A&A...183...38T} Taylor, A.~R., Seaquist, E.~R., Hollis, J.~M., et al.\ 1987, \aap, 183, 38
  
\bibitem[Tody(1993)]{1993ASPC...52..173T} 
Tody, D.\ 1993, Astronomical Data Analysis Software and Systems II, 52, 173

\bibitem[Truran \& Livio(1986)]{1986ApJ...308..721T} Truran, J.~W. \& Livio, M.\ 1986, \apj, 308, 721. doi:10.1086/164544

\bibitem[Weidemann(2000)]{2000A&A...363..647W} Weidemann, V.\ 2000, \aap, 363, 647

\bibitem[Woudt et al.(2009)]{2009ApJ...706..738W} Woudt, P.~A., Steeghs, D., Karovska, M., et al.\ 2009, \apj, 706, 738. doi:10.1088/0004-637X/706/1/738

\end{thebibliography}




\appendix



\bsp	
\label{lastpage}
\end{document}